\documentstyle[twocolumn,aps,epsf]{revtex}

\begin{document}

\draft

\title{Loading a continuous-wave atom laser by optical pumping
  techniques}

\author{Satyan Bhongale and Murray Holland}

\address{JILA and Department of Physics, University of Colorado,
  Boulder CO 80309-0440, USA.}

\date{June 23, 2000}

\wideabs{

\maketitle

\begin{abstract}
  Demonstrating that despite loss processes, Bose-Einstein condensates
  can be formed in steady state is a prerequisite for obtaining a
  coherent beam of atoms in a continuous-wave atom laser. In this
  paper we propose a method for loading atoms into the thermal
  component of a Bose condensed cloud confined in a magnetic trap.
  This method is aimed at allowing steady state dynamics to be
  achieved. The proposed scheme involves loading atoms into the
  conservative magnetic potential using the spontaneous emission of
  photons. We show that the probability for the reabsorption of these
  photons may be small .
\end{abstract}


}

\section{Introduction}

An atom laser is a device which generates an atomic beam which is both
intense and
coherent~\cite{holland,martins,spreeuw,ballagh,steck,guzman,wiseman,%
  moy,collett,kett}. It is typically defined by direct analogy with
the optical laser.  As in the case of an optical laser, if a threshold
condition is satisfied the atom laser will operate far from
equilibrium in a dynamical steady state~\cite{williams}.  In this
regime, the macroscopically occupied quantum state is continuously
depleted by loss to the output field, and continuously replenished by
pumping from an active medium. The threshold condition for the atom
laser is analogous to the critical point associated with the phase
transition from a normal gas to a Bose-Einstein condensate (BEC). A
version of the ``pulsed'' atom laser~\cite{mewes,andrews} was
demonstrated soon after the first experimental realizations of BEC in
dilute alkali gases~\cite{anderson,davis,bradley,cbradley}.

The three basic components of an atom laser are: a cavity (or
resonator), an active medium (or reservoir), and an output coupler.
The cavity is typically a three-dimensional magnetic trap which for
containment uses the force on the atomic magnetic dipole via an
inhomogeneous potential.  The active medium may be composed of a
saturated dilute thermal gas formed through evaporative cooling or
other techniques. Saturated in this context implies that the energy
distribution function of the atoms is such that thermodynamic
relaxation of the gas through binary collisions would result in the
stimulated accumulation of atoms into the ground state.  The output
coupler can be implemented by either applying a short radio-frequency
(RF) pulse to the condensate and thereby separating the condensate
into a trapped component and an un-trapped
component~\cite{mewes,andrews,bloch,lubkin}, or alternatively by
creating a situation in which the atoms tunnel out through a potential
barrier~\cite{andersonscience}.  A recent demonstration of a
quasi-continuous atom laser used an optical Raman pulse to drive
transitions between trapped and un-trapped magnetic sublevels giving
the output-coupled BEC fraction a non-zero momentum~\cite{hagley}.

To date, no continuous wave (CW) atom laser has been demonstrated in
which there is replenishing of the reservoir in direct analogy with
CW-optical laser. Replenishing the reservoir is required in steady
state to compensate for the loss due to output coupling as well as to
compensate for various intrinsic loss mechanisms such as collisions
with hot atoms from the background vapor and inelastic two-body and
three-body collisions between trapped atoms~\cite{holland,kett,hope}.

In this paper we propose a scheme for loading the thermal cloud into a
magnetic trap by optically pumping atoms from an external cold atomic
beam source. Optical pumping is necessary since the magnetic potential
is conservative and therefore a dissipative process is required for
confinement of the atoms to be achieved. When combined with
evaporative cooling, our approach may allow both steady state BEC and
the CW atom laser to be demonstrated. The crucial point is that while
one is trying to obtain a steady state BEC by injecting atoms into the
magnetic trap, the loading mechanism itself should not heat the system
or destroy the delicate condensate component.

There have been related proposals to ours using single atom ``dark
states'' for reaching low temperatures by purely optical means---even
well below the photon recoil limit~\cite{boiron,lee,leduc}. As shown
by Castin {\em et al.}\ in Ref.~\cite{castin}, the main hurdle in
achieving BEC in all these schemes is that of photon reabsorption.
Each reabsorption can remove one atom from the dark state and
increment the energy of the cloud on average by one recoil energy. We
engineer the analogous effect in our scheme to be negligible by
configuring the frequency of the emitted photons and adjusting the
spatial geometry of the system to reduce the probability of
reabsorption by the condensate.

\section{The Model}

\begin{figure}
\begin{center}\
\epsfysize=25mm
\epsfbox{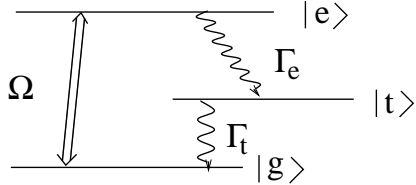}
\end{center}
\caption{
  Level scheme illustrating: the ground state $|g \rangle$, excited
  state $|e \rangle$, and trap state $|t \rangle$. A laser is used to
  couple states $|g \rangle$ and $|e \rangle$ with an intensity
  characterized by the Rabi frequency $\Omega$. Spontaneous emission
  of photons occurs with rate $\Gamma_e$ from $|e \rangle$ to $|t
  \rangle$ and rate $\Gamma_t$ from $|t \rangle$ to $|g \rangle$.}
\label{level}
\end{figure}

We consider atoms whose internal degrees of freedom are described by a
three level atomic system as shown in Fig.~\ref{level}. Our
description of the loading scheme is motivated by the experimental
setup in Ref.~\cite{cornell}. There the atoms were loaded from a
background vapor and precooled in a magneto-optical trap (MOT). They
were then launched into the
second MOT, optically pumped to the desired internal atomic
hyperfine state, and transfered to a magnetically confining potential
in which evaporative cooling was implemented. Here we study the
possibility for loading the thermal cloud into the magnetic trap
directly and continuously. This would allow the evaporative cooling to be carried
out in steady state, which is the situation considered in
Ref.~\cite{williams}.

\begin{figure}
\begin{center}\
\epsfysize=60mm
\epsfbox{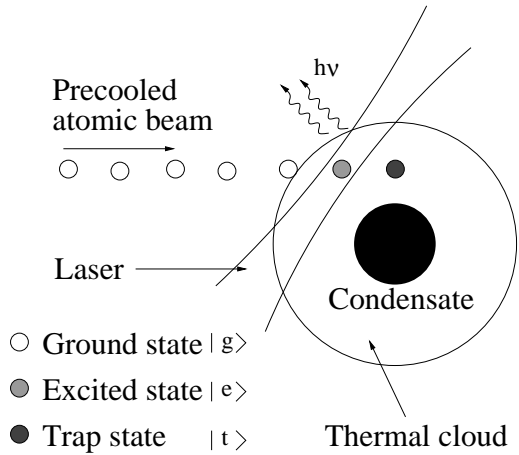}
\end{center}
\caption{ 
  Schematic diagram showing the proposed loading scheme for
  obtaining a steady state BEC and CW atom laser. Atoms entering the
  laser field are in the ground state. As the
  atoms pass through the laser field, the population is continuously
  transferred to the trap state by a sequence of coherent pumping and
  spontaneous emission.}
\label{model}
\end{figure}

Our scheme is illustrated in Fig.~\ref{model}. Atoms are injected into
the trap in state $|g\rangle$.
As the atoms enter the spatial region containing the thermal cloud
they pass through a laser field and are coherently pumped to the
excited state $|e\rangle$ from which they may spontaneously emit a
photon and end up in the trap state $|t\rangle$.

As will be shown later the spontaneously emitted photons may be out of
resonance with the bare $| t \rangle \rightarrow |e \rangle$
transition and therefore reabsorption by atoms in the thermal cloud is
greatly reduced (the exception to this general statement is the
negligible fraction of thermal atoms in the localized spatial region
of the focussed laser beam.) Using this approach atoms may be
continuously loaded into the thermal cloud and be confined by the
trapping potential without prohibitive reabsorption of the spontaneous
photons emitted in the required dissipative process.

We use a master equation~\cite{gardiner,walls} to study the evolution
of this system. We define the spontaneous emission rates between the
excited state and trap state and between the trap state and ground
state as $\Gamma_e$ and $\Gamma_t$ respectively, and for simplicity
neglect spontaneous emission directly from $|e\rangle$ to $|g\rangle$.
Assuming that the laser is on resonance with the atomic transition
frequency (detuning $\delta = 0$), the coherent interaction can be
modeled by the Hamiltonian in the interaction picture
\begin{equation}
  H =  \frac{\Omega}{2}\hbar (\sigma_+^{ge} + \sigma_-^{ge})
  \label{hamiltonian}
\end{equation}
where $\sigma_+^{ge},\sigma_-^{ge}$ are the raising and lowering
operators respectively between the ground state $|g\rangle$ and the
excited state $|e\rangle$.  The master equation for the reduced
density operator of the atom includes terms due to both the
spontaneous emission and a part corresponding to the coherent driving
source. In the interaction picture this takes the form
\begin{eqnarray}
  \frac{\partial \rho}{\partial t}& =&
  -i\frac{\Omega}{2}[\sigma_+^{ge}+\sigma_-^{ge},\rho] \nonumber \\
  &&\quad{}+\frac{\Gamma_e}{2}(2 \sigma_-^{et}\rho \sigma_+^{et}-
  \sigma_+^{et} \sigma_-^{et}\rho-
  \rho \sigma_+^{et} \sigma_-^{et}) \nonumber \\
  &&\quad{}+ \frac{\Gamma_t}{2}(2 \sigma_-^{gt}\rho \sigma_+^{gt}-
  \sigma_+^{gt} \sigma_-^{gt}\rho- \rho \sigma_+^{gt} \sigma_-^{gt}),
\end{eqnarray}
where $\sigma_+^{et},\sigma_-^{et}$ are the raising and lowering
operators respectively between the excited and trap state and
$\sigma_+^{gt},\sigma_-^{gt}$ are the raising and lowering operators
respectively between the trap state and ground state. From the above
master equation we obtain the coupled differential equations for the
evolution of matrix elements of $\rho$ with the constraint Tr$[\rho
]=1$.  The atomic population in the three levels given by the matrix
elements $\rho_{gg}(t),\rho_{ee}(t)$ and $\rho_{tt}(t)$ are obtained
for values of $\Gamma_e= 1$ and $\Omega = 5$ and initial condition
given by $\rho_{gg}(0)=1$, $\rho_{ee}(0)=\rho_{tt}(0)=0$. We consider
two different values for $\Gamma_t$; $\Gamma_t = 2$ and $\Gamma_t =
0$, which give qualitatively distinct dynamics.  The zero value for
$\Gamma_t$ corresponds to optically pumping the atom from the ground
to trap state, and leads to complete transfer of population at long
times. The non-zero value of $\Gamma_t$ gives a steady-state solution
for the level populations in which the atom cycles indefinitely, which
will be useful later for the calculation of the spectrum of the
spontaneously emitted photons.

The steady state spectrum of spontaneously emitted photons on the
$|e\rangle \rightarrow |t\rangle$ transition is given by the Fourier
transform of the two-time correlation function
\begin{equation}
  K(t+\tau,t) = \langle\sigma_+^{et}(t+\tau) \sigma_-^{et}(t)\rangle,
\end{equation}
the differential equation for which is obtained by combining the
equation for $\rho_{tg}$ and $\rho_{te}$ above along with the quantum
regression theorem (QRT)~\cite{walls}.

\begin{figure}
\begin{center}\
\epsfysize=60mm
\epsfbox{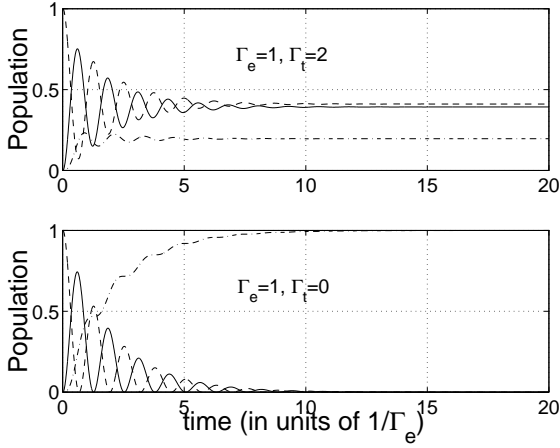}
\end{center}
\caption{
  Populations as a function of time; ground state $|g\rangle$ shown by
  a solid line, excited state $|e\rangle$ shown by dashed and trap
  state $|t\rangle$ shown by dot-dashed line. The parameters used were
  $\Omega=5, \Gamma_e=1$ and (a) $\Gamma_t=2$, (b) $\Gamma_t=0$.}
\label{pop}
\end{figure}

The QRT states that for a set of operators $Y_i$, which evolve in
steady-state according to
\begin{equation}
  \frac{\partial}{\partial \tau}\langle Y_i(\tau) \rangle = \sum_j G_{ij}
  \langle Y_j(\tau) \rangle,
\label{onetime}
\end{equation}
then two-time averages can be related to the one-time expectation
values by
\begin{equation}
  \frac{\partial}{\partial \tau}\langle Y_i(t+\tau)Y_l(t) \rangle =
  \sum_j G_{ij} \langle Y_j(t+\tau)Y_l(t) \rangle,
\label{twotime}
\end{equation}
where the $G_{ij}$'s are the same coefficients in both
Eq.~(\ref{onetime}) and Eq.~(\ref{twotime}). For a general
non-steady-state problem in which transient dynamics must be
considered, application of the QRT is typically more complicated.  For
simplicity our approach is to take a fictitious non-zero value of
$\Gamma_t$, and then consider the limit $\Gamma_t \rightarrow 0$.  The
spectrum defined as
\begin{equation}
  S(\omega) = \lim_{t \rightarrow \infty} \int_{-t}^{t}d\tau K(\tau)
  e^{-i\omega \tau}
\end{equation} 
is then given by
\begin{eqnarray}
  S(\omega)&=& \frac{\rho_{tt}^{ss} }{\pi\Omega_{\rm{eff}}} \left[
    \frac{\left(i\frac{\Gamma_e}{4}+\frac{\Omega_{\rm{eff}}}{2}
      \right) \left(\frac{\Gamma_e}{4}+\frac{\Gamma_t}{2}\right)}
    {\left(\frac{\Gamma_e}{4}+\frac{\Gamma_t}{2}\right)^2 +
      \left(\omega+\frac{\Omega_{\rm{eff}}}{2}\right)^2}
  \right] \nonumber \\
  && +\frac{\rho_{tt}^{ss} }{\pi\Omega_{\rm{eff}}} \left[
    \frac{\left(-i\frac{\Gamma_e}{4}+\frac{\Omega_{\rm{eff}}}{2}
      \right) \left(\frac{\Gamma_e}{4}+\frac{\Gamma_t}{2}\right)
      }{\left(\frac{\Gamma_e}{4}+\frac{\Gamma_t}{2}\right)^2 +
      \left(\omega-\frac{\Omega_{\rm{eff}}}{2}\right)^2} \right]
\end{eqnarray}
where $\Omega_{\rm{eff}}=\sqrt{\Omega^2-\Gamma_e^2/4}$
\begin{figure}
\begin{center}\
\epsfysize=60mm
\epsfbox{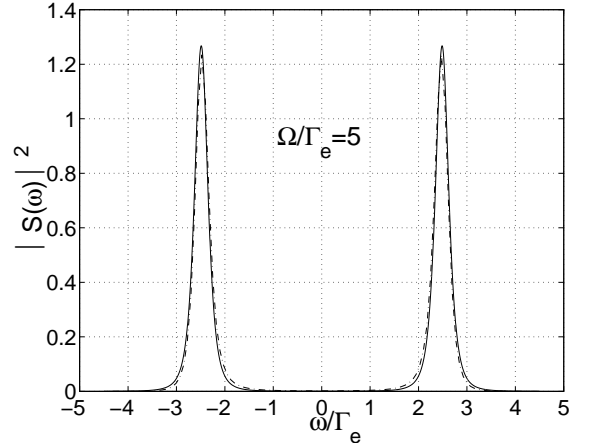}
\end{center}
\caption{ 
  Comparison of the $| e\rangle \rightarrow | t \rangle$ transition
  spectrum using the steady state calculation (solid) and the quantum
  trajectory method (dash-dot).}
\label{anal}
\end{figure}

In Fig.~\ref{anal} we plot $| S(\omega)|^2$ shown by the solid
curve for the limiting case of $\Gamma_t \rightarrow 0$. The spectrum
is AC-Stark split into two Lorentzians separated by
$\Omega_{\rm{eff}}$. This spectrum is commonly referred to as the
Autler-Townes doublet~\cite{autler,tannoudji}.

For comparison we now use the method described in Ref.~\cite{murray}
to provide an alternative calculation of the fluorescence spectrum,
extending our treatment to include the possibility for the time
dependence of $\Omega$. This involves the decomposition into ``quantum
trajectories'', where the evolution is conditional on a certain record
of dissipative events. In the case of the evolution conditional on the
spontaneous emission of no photons, the quantum trajectory
$|\psi(t)\rangle$, evolves according to
\begin{equation}
  \frac{\partial | \psi(t)\rangle}{\partial t} = \frac{1}{i\hbar}
  H_{\rm{eff}}| \psi(t)\rangle \label{traj},
\end{equation}
where $H_{\rm{eff}}$ is the non-Hermitian Hamiltonian defined by
\begin{equation}
  H_{\rm{eff}}= H - i\frac{\hbar\Gamma_e}{2} \sigma_+^{et}
  \sigma_-^{et}.
\end{equation}
For the evolution conditional on the spontaneous emission of one
photon, each quantum trajectory $|\psi_{\omega_j}(t)\rangle$ is
labelled by the frequency of the photon $\omega_j$.  If we consider a
time interval $t \in [0,\tau]$, then according to the Fourier sampling
theorem a complete description requires choosing each $\omega_j$ from
a discrete but infinite set of frequencies spaced $(2\pi/\tau)$ apart.
The observed spectrum may then be defined as
\begin{equation}
  S(\omega)= \langle \psi_{\omega}(\tau)|
  \psi_{\omega}(\tau)\rangle
\end{equation}
with (see Ref.~\cite{murray})
\begin{equation}
  \frac{d}{dt} | \psi_{\omega_j}(t)\rangle =
  \sqrt{\frac{\Gamma_e}{\tau}} \sigma_-^{et}|\psi(t)\rangle +
  \frac{1}{i\hbar}(H_{\rm{eff}} + \hbar \omega_j)|
  \psi_{\omega_j}(t)\rangle \label{spec}.
\end{equation}
By solving the coupled Eqns.~(\ref{traj}) and~(\ref{spec}) we get
\begin{equation}
\begin{array}{lcl}
  | \psi_{\omega_j}(t)\rangle & = & \left(\begin{array}{c}
      0 \\ 0 \\
      f_{\omega_j}(t)\end{array}\right)
\end{array}
\end{equation}
where
\begin{eqnarray}
  f_{\omega_j}(t) &=&q{\cal M} N^{-1}\left((i\omega_j-\Gamma_e/4)^2 +
    \Omega_{\rm{eff}}^2/4\right)^{-1}\nonumber\\
  {\cal M}&=&\sin\left(\frac{\Omega_{\rm{eff}}t}{2}\right)
  e^{-\Gamma_e t/4} (i
  \omega_j-\Gamma_e/4)\nonumber\\
  &&-\frac{\Omega_{\rm{eff}}}{2}\cos\left(
    \frac{\Omega_{\rm{eff}}t}{2}\right) e^{-\Gamma_e t/4}
  +\frac{\Omega_{\rm{eff}}}{2} e^{-i\omega_j
    t}\nonumber\\
  q&=& \frac{4}{\Omega}\sqrt{\frac{\Gamma_e}{\tau}}
  \left[\frac{\lambda_+\lambda_-}{\lambda_+ - \lambda_-}\right]
\end{eqnarray}
The single-photon spectrum which is so obtained, $| S(\omega)|^2$, is
shown by dot-dashed curve in Fig.~\ref{anal}. The normalization
constant $N$ is introduced so that $|f_{\omega_j}|^2 \rightarrow 1$ as $t
\rightarrow \infty$. We see a close agreement between the spectra
calculated by the two methods, the quantum regression theorem and the
quantum trajectory approach of solving conditional dynamics.  The
crucial observation is the absence of a significant contribution in
the zero-frequency region of the fluorescence spectrum where there
would be strong photon reabsorption by the thermal cloud. Virtually
all of the spontaneous photons are off resonance with the bare $| t
\rangle \rightarrow | e \rangle$ transition which is the resonant
frequency for all thermal atoms out of the laser field focus.

\section{
  Effect of time dependent $\Omega$ and finite detuning on the
  transient spectrum}

The assumed time independent nature of the interaction picture
Hamiltonian (\ref{hamiltonian}) may not be a realistic situation. In
an actual experiment, as the atoms pass through the laser beam they
would see a continuously changing intensity and hence a changing
$\Omega$. At the same time, existence of finite detuning $\delta$
(assumed previously to be zero for simplicity), may result in the time
dependence of the operators $\sigma_+^{ge}$ and $\sigma_-^{ge}$ given
by
\begin{equation}
\sigma_{\pm}^{ge}(t) = \sigma_{\pm}^{ge} e^{\pm i\delta t}
\end{equation}
Thus, in a more realistic situation the interaction picture Hamiltonian of
Eq.(\ref{hamiltonian}) should be replaced by the time dependent form
\begin{equation}
H(t) = \frac{\Omega(t)}{2}\hbar(\sigma_+^{ge}e^{i\delta t}+\sigma_-^{ge}e^{-i\delta t})
\end{equation}

\begin{figure}
\begin{center}\
\epsfysize=55mm
\epsfbox{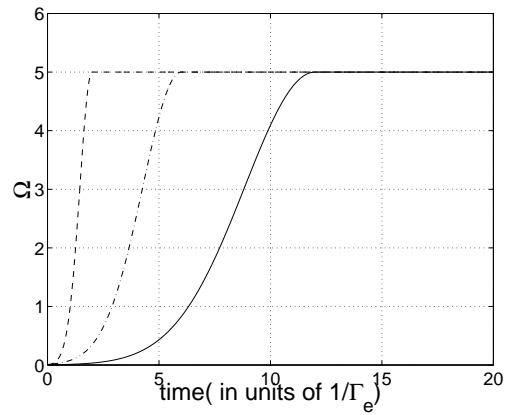}
\end{center}
\caption{ 
  $\Omega$ as a function of time, $t$. The three different curves correspond to different time scales over which $\Omega$ reaches its maximum value.}
\label{omega}
\end{figure}

\begin{figure}
\begin{center}\
\epsfysize=60mm
\epsfbox{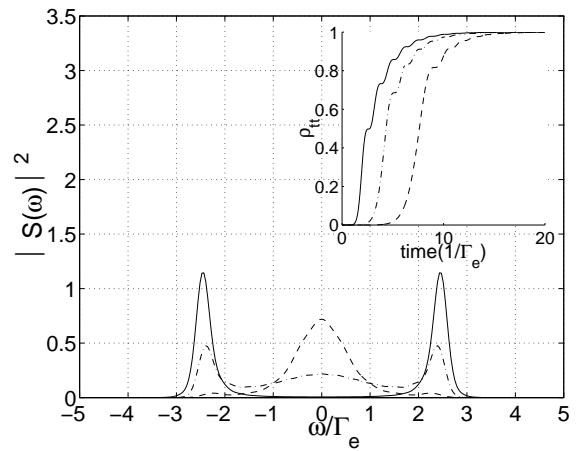}
\end{center}
\caption{ 
  Transient spectrum for $| e\rangle \rightarrow | t \rangle$
  transition for the different scenarios for $\Omega(t)$ shown in Fig.\ref{omega}.}
\label{spectrum}
\end{figure}

Using this interaction Hamiltonian we now recalculate the AC-Stark
split spectrum. Firstly we consider the $\delta=0$ case with the time
dependence solely due to $\Omega(t)$. Since the problem is a transient
one in which population is completely transferred to the trap state
over a certain time scale, the variation of $\Omega$ over this time
scale is crucial. Also, from the calculations in the previous section
it is clear that a large value of $\Omega$ is favorable for our
scheme.  We therefore consider various time scales over which $\Omega$
reaches its maximum value compared to the optical pumping time as
shown in Fig.~\ref{omega}. The corresponding spectrum shown in
Fig.~\ref{spectrum} is calculated using the method of conditional
dynamics discussed in the previous section. The calculation shows that
the crucial requirement for our model i.e.\ that the emitted photons
cannot be resonantly absorbed by the thermal cloud outside the laser
focus,is only satisfied if the atom sees a field increasing in
intensity on a time scale which is short compared to the time it takes
to optically pump the population from $|g\rangle$ to $|t
\rangle$. For a particular average velocity of the atoms in the atom
beam source, this gives the requirement for the spatial variation of
the laser intensity associated with the beam waist. Alternatively, the
laser field could be pulsed on and off to achieve a sufficiently rapid
rise time.

\begin{figure}
\begin{center}\
\epsfysize=60mm
\epsfbox{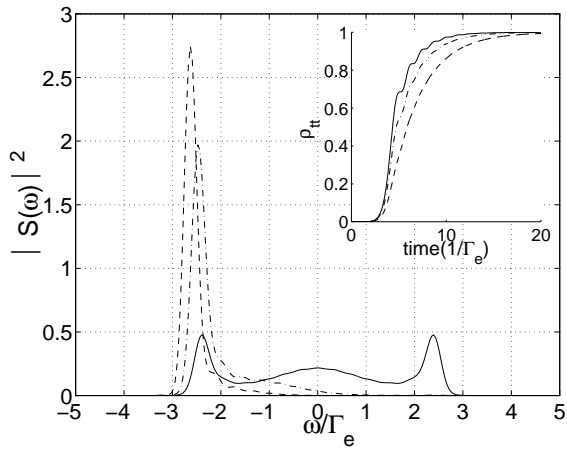}
\end{center}
\caption{ 
  Figure shows the effect of detuning on the transient spectrum for $|
  e\rangle \rightarrow | t \rangle$ transition. Dashed for $\delta =
  0$, solid for $\delta = 1$ and dot-dashed for $\delta = 2$.}
\label{detuned}
\end{figure}

Now we consider the effect of finite detuning, $\delta$, in addition
to the time dependence of $\Omega$. For the case of constant $\Omega$,
a finite $\delta$ has the effect of transforming $\Omega$ to
$\sqrt{\Omega^2+\delta^2}$~\cite{tannoudji}. Therefore, even if
$\Omega$ depends on time, it is intuitive to expect a similar effect
resulting in an increased AC-Stark splitting.  This is illustrated in
Fig.~\ref{detuned} where the spectrum is calculated for the case of
$\Omega(t)$ shown by dot-dashed curve of Fig.~\ref{omega} for various
values of $\delta$. Thus a large detuning may seem to be favorable for
our purpose since the spontaneous photons are emitted further away
from resonance. On the other hand, large detuning means slower optical
pumping rate and hence a wider laser focus will be required for the
same $\Omega(t)$ as illustrated in the inset of Fig.~\ref{detuned}.
Therefore, it is important to achieve an optimum set of parameters
that would allow a maximum possible splitting depending on the average
atomic speed and line width of the atomic levels under consideration.

\section{Conclusion}

The AC-Stark splitting of the spectrum indicates that this scheme may
potentially be used to provide a method for continuous loading of the
thermal cloud in an atom laser. In the case of constant $\Omega$ the
spectrum of the spontaneously emitted photons consists of two peaks
separated by $\Omega_{\rm{eff}}$.  Even for the realistic case of time
varying $\Omega$, large splitting between the two peaks can be
achieved provided that the rise time of the laser field intensity is
sufficiently small. The splitting can be further enhanced by
introducing a finite detuning. 

Large splitting of the spectrum is essential to avoid heating due to inelastic multiple scattering of photons. Usually heating in an optical pumping scheme is caused by photon reabsorption and re-emission. Each reabsorption can remove one atom from the tr
ap state and increment the energy of the cloud on average by one recoil energy. This effect is significantly reduced in our model since the 
spontaneously emitted photons are
off-resonance with the $|t \rangle \rightarrow | e \rangle$
transition and cannot be reabsorbed by atoms outside the laser focus. Therefore the heating of the thermal component due to photon reabsorption may be suppressed. One can also reduce the cross section for photon reabsorption by designing the
shape of the trap, for example a cigar shape or disc shape, such that
there is a limited solid angle for photon reabsorption.  Another
possibility is to align the optical pumping laser and the geometry of
the trap to use the intrinsic dipole radiation pattern to reduce
spontaneous emission into unfavorable directions.

Our model requires that the atoms be in the trap state before they are
out of the laser field, which may place a limit on the maximum speed
of atoms in the atomic beam source. However, the faster the incoming
atoms, the smaller the rise time of $\Omega$ which as previously
discussed is required to avoid the emission of photons at
zero-frequency.  The optimum speed of incoming atoms is therefore a
balance between these two requirements. Since our goal is to obtain
BEC in a steady state and not just BEC, one possibility to obtain
sufficiently slow speeds is to use a condensate prepared in another
trap as a source instead of just laser precooled atoms formed from a
magneto-optical-trap.

For a typical alkali atom, the Zeeman level diagram is composed of
many hyperfine levels. Even though our three level system is a
simplification of a real situation, most of the omitted processes,
such as imperfect optical pumping to the trapping state, will lead to
additional loss of atoms and modify the effective incident flux. This
will impact the possibility for achieving a sufficiently high pumping
rate of cold atoms to allow the CW atom laser and steady state BEC to
be achieved, as discussed in Ref.~\cite{williams}.

We would like to thank Eric Cornell, John Cooper and R. Walser for
discussions. We acknowledge support from the U.S.
Department of Energy, Office of Basic Energy Sciences via the Chemical
Sciences, Geosciences and Biosciences Division.


\begin{references}
  
\bibitem{holland} M. Holland, K. Burnett, C. Gardiner, J. I. Cirac, and
  P.  Zoller, Phys. Rev. A {\bf 54}, R1757 (1996).
  
\bibitem{martins} H. Wiseman, A. Martins, and D. Walls, Quantum.
  Semiclassic. Opt.{\bf 8}, 737 (1996).
  
\bibitem{spreeuw} R. J. C. Spreeuw, T. Pfau, U. Janicke and M. Wilkens,
  Europhysics Letters {\bf 32}, 469 (1995).  
  
\bibitem{ballagh} R. J.  Ballagh, K. Burnett, and T. F. Scott, Phys.
  Rev.  Lett. {\bf 78}, 1607 (1997).
  
\bibitem{steck} H. Steck, M.  Naraschewski, and H.  Wallis, Phys. Rev.
  Lett.  {\bf 80}, 1 (1998).  
  
\bibitem{guzman} A. M.  Guzman, M. Moore, and P. Meystre, Phys. Rev. A
  {\bf 53}, 977 (1996).  
  
\bibitem{wiseman} H. M. Wiseman, Phys Rev. A {\bf 56}, 2068 (1997).

\bibitem{moy}G. M. Moy, J. J. Hope and C. M.  Savage, Phys Rev. A {\bf
    55}, 3631 (1997).  
  
\bibitem{collett} H. M.  Wiseman and M. J.  Collett, Physics Lett. A
  {\bf 202}, 246 (1995).  
  
\bibitem{kett} W.  Ketterle and Hans-Joachim Miesner, Phys.  Rev. A
  {\bf 56}, 3291 (1997).  
  
\bibitem{williams} J. Williams, R. Walser, C.  Wieman, J. Cooper, and
  M.  Holland, Phys. Rev. A {\bf 57}, 2030 (1998).  
  
\bibitem{mewes} M. -O. Mewes {\sl et al.}, Phys. Rev. Lett.  {\bf 78},
  582 (1997).  
  
\bibitem{andrews} M.  R.  Andrews, C. G.  Townsend, H.  -J. Miesner, D.
  S. Durfee, D. M.  Kurn, and W.  Ketterle, Science {\bf 275}, 637
  (1997).  
  
\bibitem{anderson} M. H.  Anderson, J. R.  Ensher, M. R.  Matthews, C.
  E.  Wieman, and E. A.  Cornell, Science {\bf 269}, 198 (1995).
  
\bibitem{davis} K. B. Davis, M.- O Mewes, M.  R. Andrews, N.  J. van
  Druten, D. S. Durfee, D. M.  Kurn, and W.  Ketterle, Phys.  Rev.
  Lett.  {\bf 75}, 3969 (1995).
  
\bibitem{bradley}C. C. Bradley, C. A.  Sackett, J. J. Tollett, and R.
  G.  Hulet, Phys. Rev. Lett {\bf 75}, 1687 (1995).
  
\bibitem{cbradley}C. C. Bradley, C. A.  Sackett, and R.  G.  Hulet,
  Phys. Rev. Lett {\bf 78}, 985 (1997).  
  
\bibitem{bloch}I.  Bloch, T.  W. Hansch, and T. Esslinger, Phys. Rev.
  Lett.  {\bf 82}, 3008 (1999).  
  
\bibitem{lubkin} G. B. Lubkin, Physics Today, {\bf 52}, 17 (1999).
  
\bibitem{andersonscience} B. P. Anderson and M. A. Kasevich, Science
  {\bf 282}, 1686 (1998).

\bibitem{hagley}E. W. Hagley {\sl et al.}, Science {\bf 283}, 1706
  (1999).  
  
\bibitem{hope}J. J. Hope, Phys. Rev. A {\bf 55}, R2531 (1997).
  
\bibitem{boiron}D.  Boiron, A.  Michaud, P. Lemonde, and C. Salomon,
  Phys.  Rev. A {\bf 53}, R3734 (1996).  
  
\bibitem{lee}H.  J. Lee, C.  S.  Adams, M.  Kasevich, and S. Chu,
  Phys. Rev.  Lett.  {\bf 76}, 2658 (1996).  
  
\bibitem{leduc}J.  Lawall, S. Kulin, B.Saubamea, N.  Bigelow, M.
  Leduc, and C.  Cohen-Tannoudji, Phys.  Rev.  Lett. {\bf 75}, 4194
  (1995).  
  
\bibitem{castin} Y. Castin, J.  Cirac, and M.  Lewenstein, Phys. Rev.
  Lett.  {\bf 80}, 5305 (1998).  
  
\bibitem{cornell}E. A.  Cornell, C. Monroe, and C. E.  Wieman, Phys.
  Rev.  Lett. {\bf 67}, 2439 (1991).
  
\bibitem{gardiner}C. W.  Gardiner, {\sl Quantum Noise} (Springer,
  Berlin, 1991).  
  
\bibitem{walls}D. F.  Walls and G. J. Milburn, {\sl Quantum Optics},
  (Springer, Berlin, 1994).  
  
\bibitem{murray} M.  Holland, Phys. Rev.  Lett. {\bf 33}, 5117 (1998).

\bibitem{autler}S. H.  Autler and C. H. Townes, Phys. Rev. {\bf 100}, 703 (1955). 
  
\bibitem{tannoudji}Claude Cohen-Tannoudji,Jacques Dupont-Roc, and G.
  Grynberg, {\sl Atom--Photon Interactions} (John Wiley and Sons, New
  York, 1992).

\end{references}
\end{document}